\documentclass[iop]{emulateapj}

\usepackage{multirow}
\usepackage{natbib}
\usepackage{longtable,rotate}

\newcommand{\noprint}[1]{}

\shorttitle{}
\shortauthors{M. S. Shaw et al.} 
		
\slugcomment{Accepted in The Astronomical Journal.}

\usepackage{lscape}

\begin{document}
\title{Photometrically Triggered Keck Spectroscopy of {\it Fermi} BL~Lac Objects}
\setlength{\paperheight}{11.0in}
\setlength{\paperwidth}{8.5in}

\author{Michael S. Shaw\altaffilmark{1}, Alexei V. Filippenko\altaffilmark{2},  Roger W. Romani\altaffilmark{1}, S. Bradley Cenko\altaffilmark{2,3}, and Weidong Li\altaffilmark{2,4}
}

\altaffiltext{1}{Department of Physics/KIPAC, Stanford University, Stanford, CA 94305}
\altaffiltext{2}{Department of Astronomy, University of California, Berkeley, CA 94720-3411}
\altaffiltext{3}{Astrophysics Science Division, NASA Goddard Space Flight Center, Mail Code 661, Greenbelt, MD, 20771, USA}
\altaffiltext{4}{Deceased 12 December 2011.}


\begin{abstract}
We report on Keck spectra of ten {\it Fermi} blazars. J0622+3326, previously 
unobserved, is shown to be a flat-spectrum radio quasar at redshift $z=1.062$.  The others are known BL~Lac-type objects
that have resisted previous attempts to secure redshifts.  Using a photometric 
monitoring campaign with the 0.76\,m Katzman Automatic Imaging Telescope at Lick Observatory, we
identified epochs when the relativistic jet emission was fainter than usual,
thus triggering the Keck spectroscopy.
This strategy gives improved sensitivity to stars and ionized gas in the host galaxy, thereby providing 
improved redshift constraints for seven of these sources.
\end{abstract}



\section{Introduction}

Blazars are the brightest extragalactic
point sources in the gamma-ray (and microwave) bands; study of their
population and evolution are key to high-energy astrophysics. Many 
blazars are so-called BL~Lacertae objects, where the optical
spectrum is dominated by continuum synchrotron radiation. These tend to be 
hard-spectrum $\gamma$-ray sources, and appear to provide a substantial fraction 
of the $E>10$\,GeV extragalactic $\gamma$-ray background \citep{1FHL}. 
However, the extreme continuum domination in the optical makes it particularly 
difficult to obtain redshifts, compromising studies of the BL~Lac population and 
its cosmic evolution \citep[][hereafter S13]{bll}.

Using two years of sky-survey data, The {\it Fermi} Second Source Catalog \citep[2FGL;][]{2FGL} reports on the 1873 most significant sources detected in the Large Area Telescope \citep[LAT;][]{atw09}. 
More than half of these objects are associated with active galactic
nuclei (AGNs) --- in particular, jet-dominated ``blazars,'' which are generally compact, flat-spectrum radio sources.
Of the 1121 such associations (1017 at $|b|>10^\circ$) 
in the Second Catalog of AGNs detected by the {\it Fermi} LAT \citep[2LAC;][]{2LAC},
410 were flagged as BL~Lacs. In S13, additional spectroscopy increased the BL~Lac
number to 475. However, despite extensive observations with 10\,m-class telescopes, secure spectroscopic redshifts
were obtained for only 209 BL~Lac objects (44\%). Fortunately, with high-quality spectra one can derive
constraints on the redshift of the other BL~Lac objects, especially using standard
assumptions about the host-galaxy luminosity. In S13, such constraints (or redshifts) were
obtained for $95\%$ of the BL~Lac objects, providing substantial improvement in our understanding
of BL~Lac evolution (Ajello et al. 2013).

	The principal barrier to spectroscopic redshifts is the extreme dominance of the
optical continuum. This continuum can, however, be highly variable; the constraints can 
improve and additional redshifts may be obtained when the source is in a ``low'' state.
In fact, \citet{fsrq} describe five LAT blazars which qualify as BL~Lac objects when in 
a high state, but show broad-line emission with equivalent width (EW) $>5$\,\AA\ in low states, 
such that they would be defined as flat-spectrum radio quasars (FSRQs) at these epochs.
We have therefore used photometry from an ongoing blazar optical monitoring campaign to
trigger low-state spectroscopy of blazars lacking secure redshifts. Here
we report on the results of those observations.

In \S \ref{sec:phot} we describe the monitoring campaign and the target selection for spectroscopy. 
Section \ref{sec:spec} presents the spectroscopic observations and reductions.
We discuss redshifts and redshift constraints in \S \ref{sec:z}. Our conclusions are summarized in \S \ref{sec:conclusion}, 
and we note how similar campaigns may provide the best prospect for additional BL~Lac redshifts.
In this paper, we assume an approximate concordance cosmology: 
$\Omega_m=0.3$, $\Omega_\Lambda=0.7$, and H$_0=70$\,km\,s$^{-1}$\,Mpc$^{-1}$.


\section{KAIT Photometry}
\label{sec:phot}

\begin{figure*}
\epsscale{1.3}
\vspace{-250pt}
\hspace*{-40pt}
\plotone{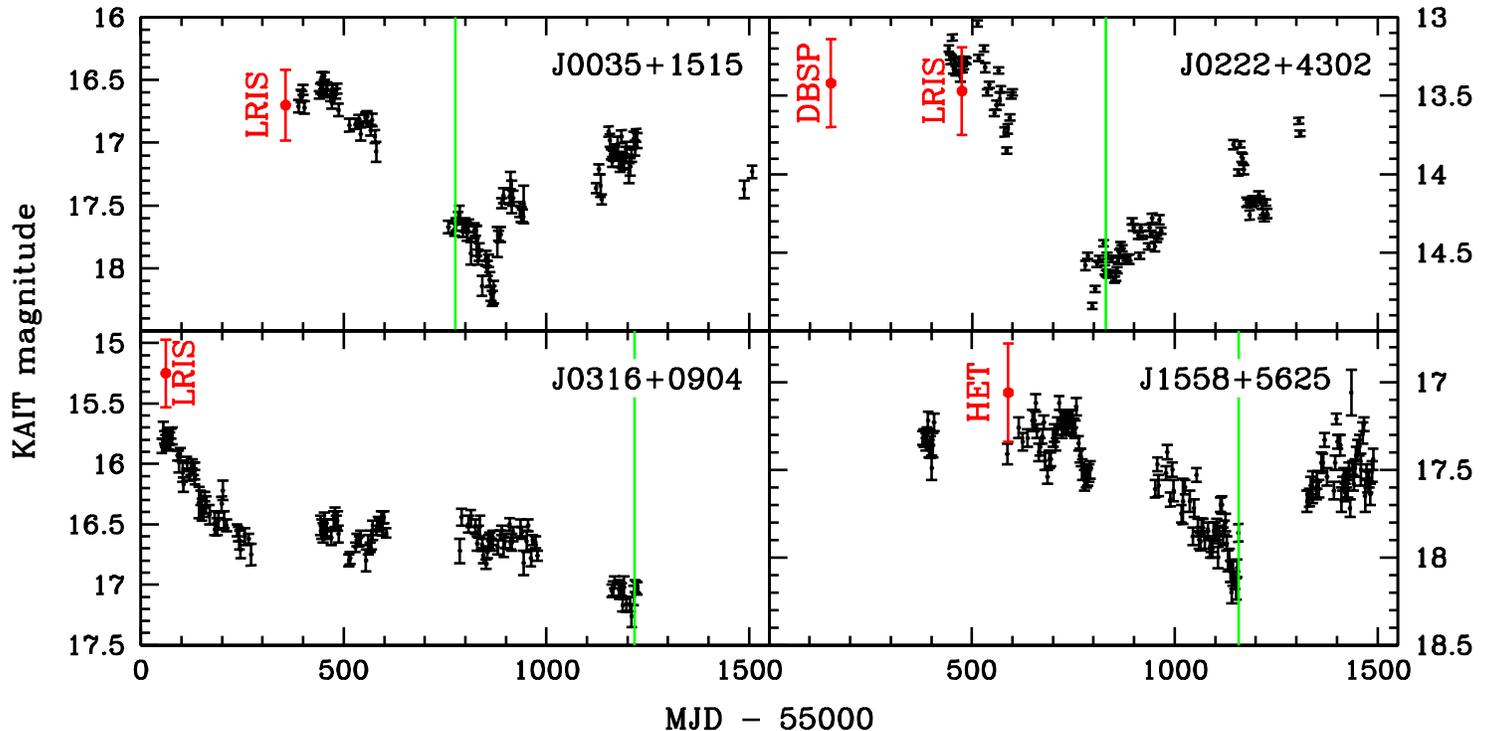}
\vspace{-40pt}
\caption{KAIT photometry (black error bars) for four monitored BL~Lac objects. Vertical green lines 
indicate the date of the triggered (i.e., low-state) spectroscopic observation, all with 
Keck LRIS. Red error bars indicate the dates and approximate fluxes of our previous spectroscopy.}
\label{fig:lightcurves}
\end{figure*}

	To have the best chance of detecting significant short-timescale variability,
we selected LAT blazars that are relatively bright in both the $\gamma$-ray
and optical bands. In the $\gamma$-ray bands, we selected sources having
a likelihood test statistic, TS $>100$ ($\gtrsim 10\sigma$; for details see \citealt{2FGL}, and references therein) in the first year of LAT observations. In the optical, the selected sources had quiescent magnitude $R<18$. After restricting to high Galactic latitude ($|b|>10^\circ$) and $-25^\circ< \delta <70^\circ$, and dropping a 
few famous sources that are well monitored by other programs, we finalized a sample of 156 LAT AGNs.

The photometric monitoring was obtained with the Katzman Automatic Imaging Telescope \citep[KAIT;][]{kait},
a 0.76\,m robotic telescope at Lick Observatory on Mt. Hamilton whose primary mission is to discover and monitor young, nearby supernovae \citep[e.g.,][]{ganeshalingam10,leaman11,li11}. The blazars were added to the queue,
with 30--60\,s white-light (unfiltered, but roughly $R$ in overall response; \citealt{li03}) exposures to be taken on a 3-day cadence, subject
to weather and angular separation from the Sun. An automated photometry pipeline extracts simple
aperture magnitudes each night, calibrated against the USNO B1.0 $R$-band magnitudes of a manually selected set of stars in each target field. Fluxes and updated light curves for newly observed targets 
are distributed for inspection the following day. 

The blazars do indeed show a wide range of variability on a variety of
time scales. A principal goal of the program is to measure optical/$\gamma$-ray correlations;
we will report on this analysis elsewhere. The monitored set does include many
BL~Lac objects lacking spectroscopic redshifts (50 at the start of the campaign), and 
we can improve the understanding of these sources with sensitive spectroscopy
at low states. As part of an ongoing program to study supernovae and other transients, we have
access to the Keck 10\,m telescopes on an approximately monthly basis.
Thus, KAIT-monitored BL~Lac objects lacking redshifts and that happened to be in an 
especially low state during a Keck run were selected as spectroscopic targets.

Figure \ref{fig:lightcurves} shows the light curves of four of these targets. For each
target we had previous observations with 5--10\,m-class telescopes, establishing the sources
as BL~Lac objects with a very low EW limit on any emission or absorption lines. The triggered KAIT epoch is 
indicated by a green line and the approximate magnitude during previous spectroscopy, scaled
relative to the flux at the KAIT-selected epoch, is shown in red. At the triggered epochs,
the objects are generally $\gtrsim 1$\,mag fainter than normal.


\section{Keck LRIS Spectroscopy}
\label{sec:spec}

Spectra were obtained with the Low Resolution Imaging Spectrometer \citep[LRIS;][]{oke95}
on the Keck-I telescope. Observations were taken through a slit of width $1''$
and the 5600~\AA\ dichroic filter was used. On the blue side, we employed 
the 600/4000 grism, for an effective resolution of $4\,$\AA, a dispersion of $0.63\,$\AA\ per pixel, and coverage from the atmospheric 
limit (3100\,\AA) to $\sim 5620$\,\AA). On the red side, the 400/8500 
grating was used, for an effective resolution of 7\,\AA, a dispersion of $1.16\,$\AA\ per pixel, and coverage from 5500\,\AA\ to 10,200\,\AA.

LRIS has an atmospheric dispersion corrector, and we also generally
oriented the slit along the parallactic angle \citep{fil82} to minimize differential losses. At least two exposures were taken of every target to allow simple removal of cosmic rays. The MJD of the exposure and the net integration times are listed in Table \ref{tab:props}. 
These are bright targets for Keck/LRIS; thus, spectra were usually taken during twilight 
or in nonphotometric conditions. Typically we are not background limited and the resulting 
spectra have a high signal-to-noise ratio (S/N) per resolution element, $\sim$ 50 to $> 300$.

	In addition to nine known BL~Lac objects selected by continuum drops, we observed 
J0622+3326, associated with 2FGL J0622.9+3326 at $> 99$\% probability. This is a $>20\sigma$ 
source in \citet{2FGL}, and one of the brightest {\it Fermi} objects having previously 
unknown optical type.

\begin{figure*}
\epsscale{1.3}
\vspace{-20pt}
\hspace*{-40pt}
\plotone{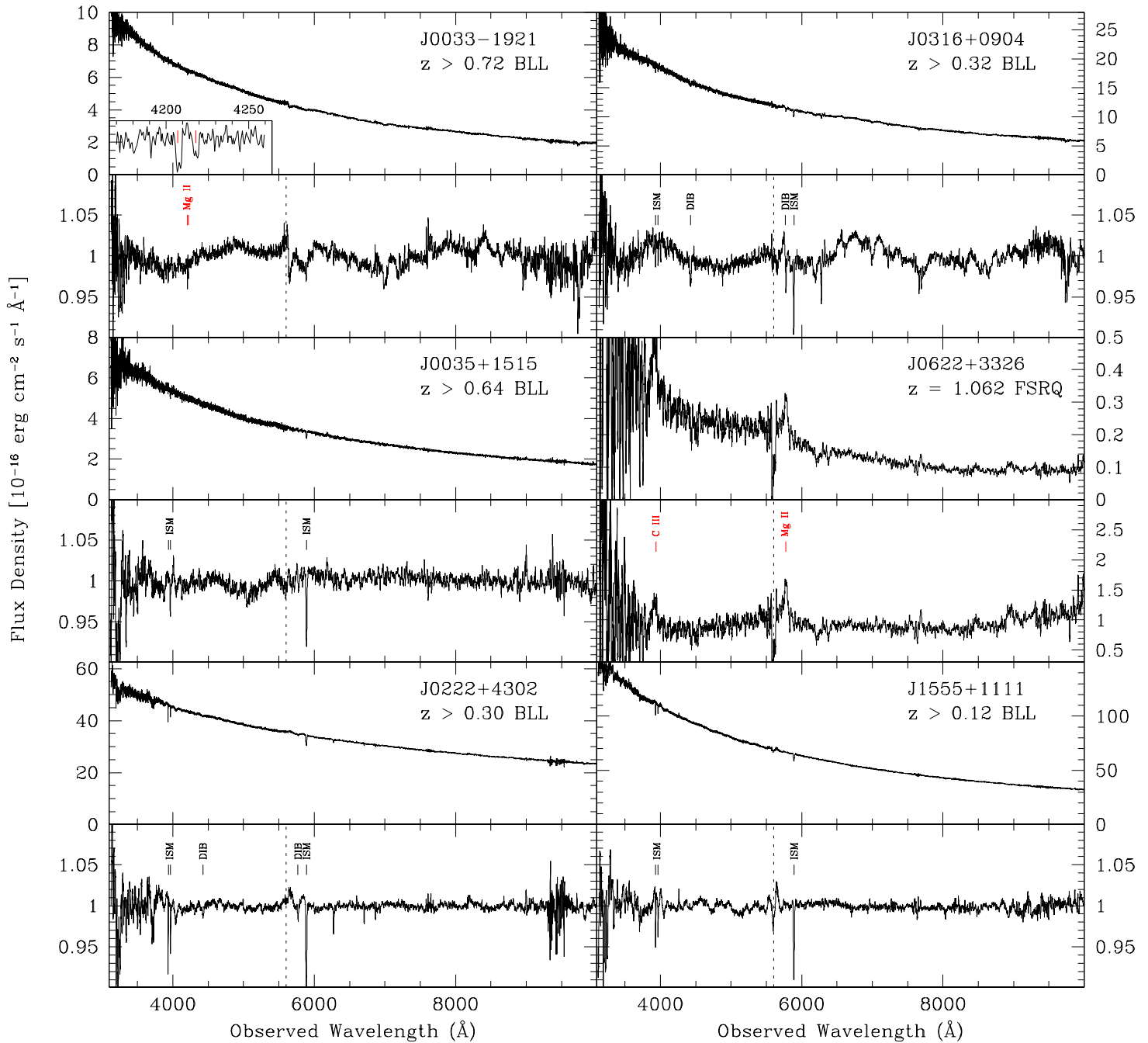}
\vspace{-40pt}
\caption{Keck/LRIS spectra of {\it Fermi} blazars.  Each object is presented twice  --- directly in the upper panel, 
and then divided by the best-fit power law (lower panel). Spectroscopic redshifts
or lower limits are based on the two lines marked in red; other lines of interest are black. If constrained by an absorption doublet, this is highlighted in an inset figure, with the lines again marked in red.
Each plot is labeled with the most constraining $z$ limit --- two digits for limits
from nondetection of host galaxies, and more than two digits for intervening absorption systems. 
See Table \ref{tab:props} for the precise spectroscopic limit. The dichroic position is marked by a dotted vertical line.}
\label{fig:spectra1}
\end{figure*}

\addtocounter{figure}{-1}
\begin{figure*}
\epsscale{1.3}
\vspace{-170pt}
\hspace*{-40pt}
\plotone{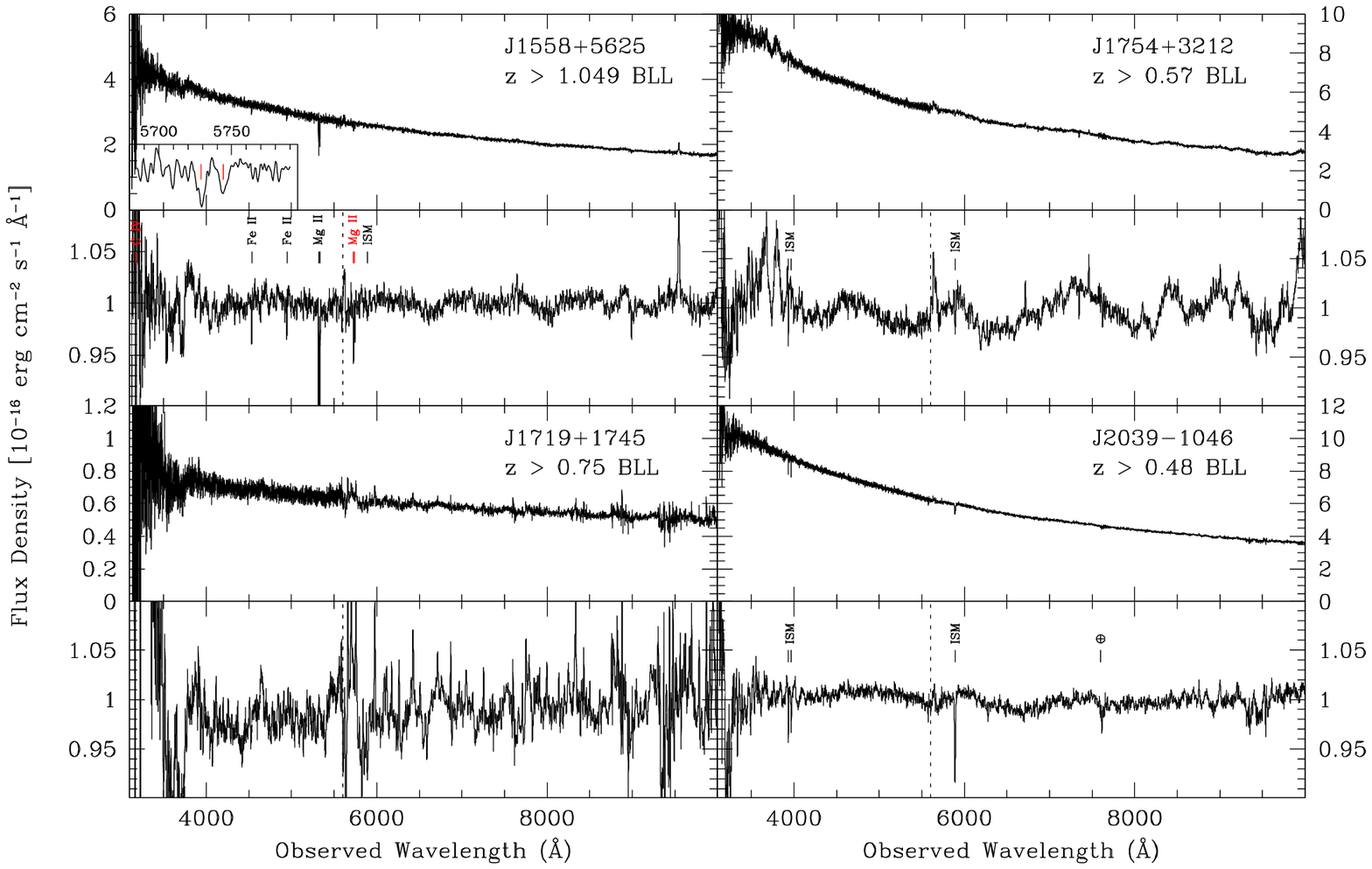}
\vspace{-40pt}
\caption{{\it Continued}}
\label{fig:spectra2}
\end{figure*}

Spectroscopic data reduction proceeded using the IRAF package \citep{tod86, val86} with 
standard techniques. Data were bias-subtracted and flat-fielded (with dome flats).
Wavelength calibration employed afternoon spectra of lamps and was confirmed by night-sky emission lines. We used an optimal extraction algorithm \citep{val92}, maximizing 
the final S/N. Spectra, especially those taken with the red-side camera (which utilizes a thick CCD chip), were visually cleaned of residual cosmic-ray contamination affecting only individual exposures.  

Spectrophotometric calibrations were made using standard stars from 
\citet{oke90} and \citet{boh07}. These were generally observed on the
same night as the BL~Lac targets, but in a few cases they were unavailable 
and standard stars from other nights were substituted. 
Since BL~Lac spectra are generally simple power laws, we used our objects to monitor residual 
errors in the sensitivity function. Nevertheless, in these very high-S/N data, some 
residual calibration errors (at the $\sim 1$--3\% level) remain. 

Spectra were corrected for atmospheric extinction using typical values
for Mauna Kea \citep{kri87}.
We used IRAF's dereddening function with the 
recently updated Galactic extinction maps \citep{sf11} to remove any Galactic reddening. We made no attempt to 
remove intrinsic (host-galaxy) reddening.

Telluric templates were generated from the standard-star observations on each 
night. We determined templates for the oxygen and water-line complexes and
corrected individually for the telluric absorptions of these two species. We find significant residuals only in high-S/N spectra taken under varying, nonphotometric atmospheric conditions. 

After these calibrations, we co-added individual exposures separately on 
the red and blue sides, weighted by S/N. Finally, we stitched the red-side and 
blue-side spectra together, and these are presented in Figure \ref{fig:spectra1}.

We estimate systematic errors in absolute 
spectrophotometry of $\sim 30$\% \citep{cgrabs}, primarily due to variable slit losses and changing atmospheric conditions; relative 
spectrophotometry is likely considerably better.
 
A few systematic artifacts are still visible in the stitched spectra (Fig. \ref{fig:spectra1}). Occasionally, 
there is a systematic step at the dichroic cutoff. Due to the strongly varying 
sensitivity through the cutoff wavelength, our calibrations are sensitive to 
very subtle perturbations (i.e., from varying temperature); thus, the flux variations near 
the dichroic are probably not real.
In J0316+0904 and J1719+1745, residual telluric features near 7600\,\AA\ are present 
in the reduced spectra. Due to the varying atmospheric conditions, the standard stars 
did not perfectly model the telluric absorption in the object spectra, and division 
left a detectable residual. We visually compared any residuals to previous spectra of
these objects; none were found to be consistent, suggesting that they are not weak, intrinsic
features. These residual effects are at the few percent level;
the BL~Lac spectra are impressive power laws, and our data allow us to search for
subtle departures associated with any underlying broad-line region and AGN host.

\section{Redshift Measurements and Constraints}
\label{sec:z}

	Since previous spectroscopic observations of these obdurate sources did not
provide redshifts (S13), even spectra in a low state require careful analysis.
Such work is motivated by the need to have high redshift completeness in the LAT
sample when probing source evolution \citep{aje12}.

	In this program we observed one $\gamma$-bright LAT blazar without previous spectroscopic
study. J0622+3326 proved to be an FSRQ, and a single 640\,s observation in twilight
sufficed to identify highly significant ($>5\sigma$), broad \ion{C}{3}] (rest 1909\,\AA) 
and \ion{Mg}{2} (rest 2795.5, 2802.7\,\AA) emission at $z=1.062$, measured by 
cross-correlation analysis with the ``rvsao'' tool \citep{rvsao}. All other sources were
known BL~Lac objects and, even in a low state, we did not find significant broad emission-line features.
More subtle constraints on redshift must be extracted.

Intervening metal-line absorption systems provide secure lower limits on the blazar
redshift.  We search for \ion{Mg}{2} $\lambda\lambda$2795.5, 2802.7 and \ion{C}{4} $\lambda\lambda$1548.2, 1550.77 
absorption lines in all spectra.  Visual inspection revealed two sources with absorption 
doublets, as reported in Table \ref{tab:props}. We fit the absorption with a two-Gaussian template having 
wavelength spacing that scales with $z$, and free amplitudes. For the doublet to be significant we require 
the stronger (weaker) component to be $\ge5\sigma$ ($\ge3\sigma$), and the doublet ratio to be in the 
acceptable range (between 1:1 and 2:1, with the blue line dominant), following \citet{nes05}.

J0033--1921 shows significant \ion{Mg}{2} absorption at $z=0.505$. This feature was too weak to detect 
in previous (5\,m-class) observations. It is, however, less constraining than the model-dependent host lower 
limits discussed below. In J1558+5625, we confirm the previously known system at $z=0.909$, and we find a 
weaker, but still significant \ion{Mg}{2} absorption doublet at $z=1.049$. This model-independent constraint 
is consistent with and stronger than the host limits, and pushes this BL~Lac out to an interesting redshift.

\begin{figure*}
\epsscale{1.45}
\vspace{-10pt}
\hspace*{-70pt}
\plotone{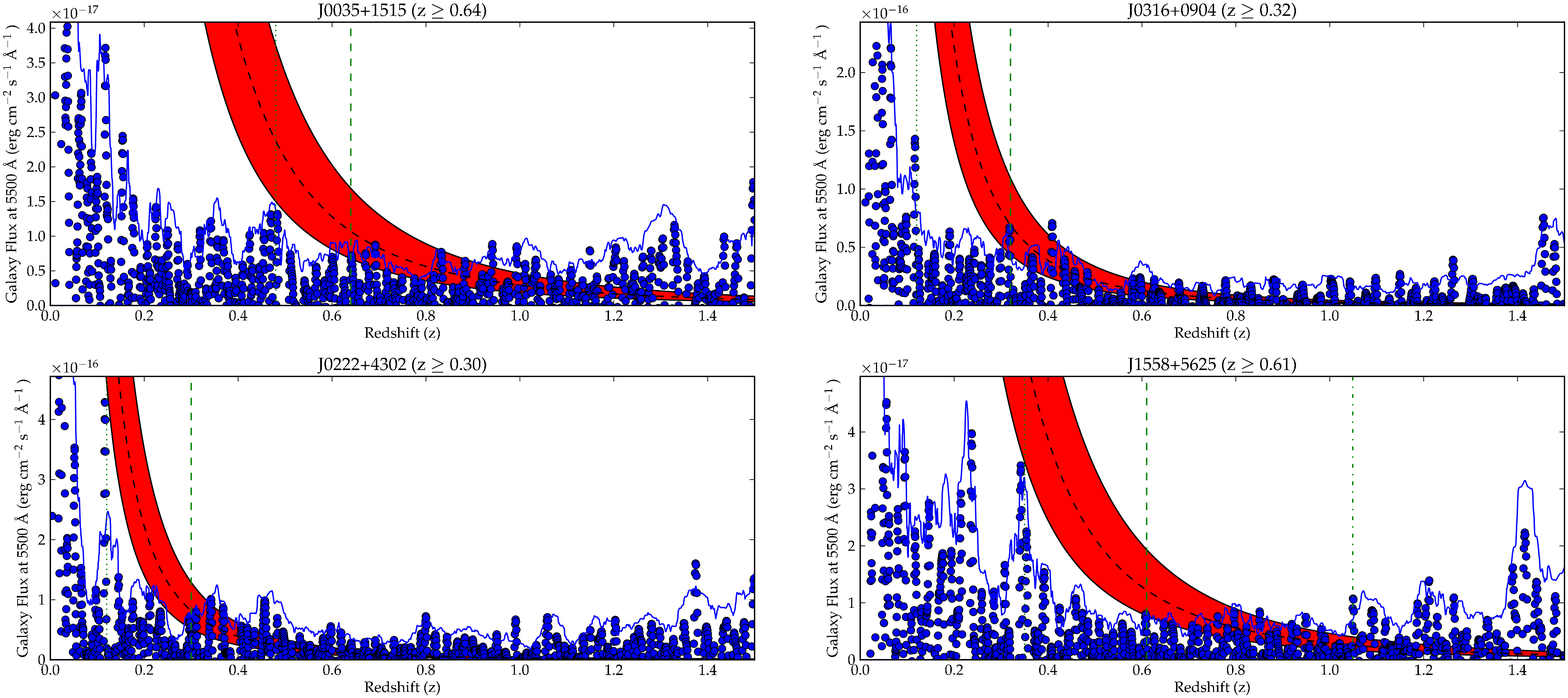}
\vspace{-20pt}
\caption{Best-fit host-galaxy fluxes as a function of $z$ (blue dots). The blue line is an estimate of the $2\sigma$ local systematic flux error. The $1\sigma$ range about the expected host flux for $M_R=-22.5 \pm 0.5$\,mag is represented by a red band.  
Vertical green dashed lines represent the $z_{\rm min}(-22.5)$ limit.  For comparison, 
dotted green lines give the previous (S13) lower limit from archival spectroscopy. In J1558+5625, 
we also have a secure lower limit from an intervening \ion{Mg}{2} system (green dash-dot line).}
\label{fig:fitter}
\end{figure*}

We next follow the technique developed by S13 to derive redshift lower limits based on nondetection 
of galaxy hosts. In brief, we fit the observed BL~Lac spectra with two components: a power law for 
the core emission, and an elliptical host-galaxy template for the galaxy emission. All spectra are 
bandpass filtered before fitting to minimize sensitivity to the residual systematic noise.
The fit thus returns power-law parameters, host flux, and error estimates at each trial redshift.
The power-law errors in Table \ref{tab:props} include systematics from flux calibration and a 
spectral-index systematic estimated by separately fitting the red and blue halves of the spectra.

BL~Lac hosts have been described as standard candles with $R = -22.9$\,mag \citep{urr00,sba05}. In our
study of the {\it Fermi} BL~Lac objects (S13) we recalibrated this magnitude, finding less luminous ($R = -22.5$\,mag) hosts
for the LAT-detected BL~Lacs. When the fitted host at a trial $z_i$ is fainter than
the standard-candle luminosity, we exclude that redshift, as shown in Figure \ref{fig:fitter}. 
To be conservative, we adopt here the $R = -22.5$\,mag
calibration, assuming $R = -22.9$\,mag gives slightly larger lower bounds. (For further details see S13.)
These BL~Lac redshift lower limits are reported in Table \ref{tab:props}; they are
typically more stringent than those quoted by S13, with the improvement given as $\Delta z_{\rm min (-22.5)}$.

We can also provide statistical upper limits on redshift from the lack of Ly$\alpha$ absorption 
features.  Given a known density of Ly$\alpha$-forest absorbers, we define a redshift range $\Delta z$
over which we are likely (probability $> 0.68$; $1\sigma$) to find an absorption feature,
\begin{equation}
\Delta z = 0.167\, e^{({\rm EW}_{\rm rest} - 0.24) / 0.267)} \times \left(\frac{3100}{1215}\right)^{-1.85},
\label{eqn:upper}
\end{equation}
where ${\rm EW}_{\rm rest}$ is the measured EW limit (S13). We find $z_{\rm max} = 1.55 + \Delta z$, 
where 1.55 is the redshift where Ly$\alpha$ reaches the blue end of our spectra. The
``proximity effect'' is negligible at these redshifts; for a derivation, see S13.
Upper limits $z_{\rm max}$ are reported for each BL~Lac object in Table \ref{tab:props}; they 
generally vary from 1.6 to 1.8 based on the S/N of the spectrum.


\section{Discussion and Conclusions}
\label{sec:conclusion}

	Most BL~Lac samples have redshifts for $<50$\% of the sources, even after extensive
10\,m-class observations. In S13 we showed, using other redshift constraints (intervening
absorbers and host flux limits), that the unsolved BL~Lac objects were generally at higher redshift 
than those having spectroscopic redshifts. Thus, redshift incompleteness introduces serious biases 
into studies of BL~Lac evolution. Although use of the redshift constraints can partially obviate this
problem (Ajello et al. 2013), it is important to identify as many redshifts as possible and/or 
improve constraints to extract the tightest limits on the source $z$.

	Several of our BL~Lacs are well-known objects, with previous redshift estimates presented 
in the literature. These estimates have propagated to the NASA Extragalactic Database (NED).
Our data do not confirm the emission or absorption lines reported for J0033--1921 \citep[$z=0.61$;][]{pir07},
J1558+5625 \citep[$z=0.30$;][]{fal98}, and J1719+1745 \citep[$z=0.137$;][]{dse05}, despite the increased
wavelength coverage and larger S/N in the present study, so we believe that those values are
erroneous.

	We have shown how triggered low-state spectroscopy can provide improved redshift constraints
on BL~Lac sources. Although none of the BL~Lac objects here showed significant broad-line emission
or host-galaxy features (and are thus highly continuum dominated even at these low states),
for seven of nine sources the analysis resulted in a smaller allowed redshift range. 
For J1555+1111, the archival (pre-KAIT) observation was actually at a quite low state,
so the new observation during a minimum of the KAIT light curve, was, in fact, at a brighter state, and thus less constraining. For J1754+3212, the short exposure under poor conditions did not match the quality of the previous spectra, so that, despite the fainter continuum state, we obtained similar redshift constraints. We also provide a spectroscopic redshift for J0622+3326, a newly identified LAT FSRQ, and power-law parameters for the BL~Lac optical continua, which can be useful to those 
modeling the sources' spectral energy distributions.

	While triggered 10\,m spectroscopy helps, other observations may also be useful in
restricting the allowed redshift range. For example, for J0222+4302 (3C~66A), \citet{fur13} have used
{\it HST} ultraviolet (UV) spectroscopy to study the Ly$\alpha$ absorbers. These show that $z>0.33447$ and
$z<0.41$ (99\% confidence). Such UV spectroscopy is particularly powerful at lowering the
$z_{\rm max}$ upper limits. UV photometry can also reveal continuum drops associated with the
onset of Ly$\alpha$ absorption. \citet{rau12} report an estimate of $z_{\rm phot}=1.28^{+0.14}_{-0.17}$
for J0035+1515 and an upper limit of $z< 1.35$ for J1555+1111. Both are consistent
with our spectroscopic limits.

	An additional constraint is available when high-quality imaging can
resolve the BL~Lac host from the point-source core, providing an estimate of the flux, and thus, through the standard-candle approximation, of the redshift \citep[e.g.,][]{sbaim,mei10}. Although
none of the present sources have imaging estimates, these can provide useful constraints
to $z>0.7$ and so might be pursued in the future. Finally, a combination of imaging
and spectroscopy, with an optimal aperture extracted from integral field unit (IFU) spectroscopic
observations, can help isolate the host for improved measurements or limits on its
spectrum. Of course, both of these techniques benefit from decreased core luminosity,
so observations triggered by low states in the light curve will give optimal results.

	Clearly, the measurement of redshifts for BL~Lac objects with extreme continuum dominance,
such as those studied here, will remain challenging for the foreseeable future. However,
the motivation to obtain complete spectroscopic identification of high-energy selected
samples is becoming increasingly strong, as these can provide improved understanding
of BL~Lac evolution and the cosmic $\gamma$-ray and optical backgrounds. As shown here,
coordination with photometric monitoring campaigns, to obtain the spectra and/or host 
imaging in the lowest state possible, is important as it makes optimal use of expensive
10\,m-class telescope time.
	
\acknowledgements

The work was supported in part by NASA grants NNX10AU09G, GO-31089,
NNX12AF12GA, and NAS5-00147.  A.V.F. and his group at U.C. Berkeley
are also grateful for support from Gary and Cynthia Bengier, the
Richard and Rhoda Goldman Fund, the Christopher R. Redlich Fund, the
TABASGO Foundation, and NSF grant AST-1211916.  KAIT and its ongoing
operation were made possible by donations from Sun Microsystems, Inc.,
the Hewlett-Packard Company, AutoScope Corporation, Lick Observatory,
the NSF, the University of California, the Sylvia and Jim Katzman
Foundation, and the TABASGO Foundation.  Some of the data presented
herein were obtained at the W. M. Keck Observatory, which is operated
as a scientific partnership among the California Institute of
Technology, the University of California, and NASA. The Observatory
was made possible by the generous financial support of the W. M. Keck
Foundation.  We dedicate this paper to the memory of our dear friend
and collaborator, Weidong Li, whose unfailing devotion to KAIT made
this work possible; his premature, tragic passing has deeply saddened
us.

{\it Facilities:} \facility{Fermi}, \facility{Lick:KAIT}, \facility{Keck:I (LRIS)}.

\bibliography{references}

\clearpage
\begin{landscape}
\begin{deluxetable*}{ccrlcccccrcrcrr}
\tablecolumns{15}
\tabletypesize{\tiny}
\tablecaption{BL~Lac Spectral Properties}
\tablehead{
\colhead{2FGL} & \colhead{RA~($^{h\,m\,s}$)} & \colhead{Dec.~($^{\circ\,' \,''}$)} & \colhead{Common Name} & \multicolumn{3}{c}{Name\ \ \ \  \ \ \ $\log F_{\nu,10^{14.7}}^\tablenotemark{a}$\ \ \ \ \ \ \ \ \ \ $\alpha$\ \ \ \ \ \ } & \colhead{$z_{\rm spec}$} & \colhead{$z_{\rm min}$} & \colhead{$\Delta z_{\rm min}$} & \colhead{$z_{\rm max}$} & \colhead{Type$^\tablenotemark{b}$} & \colhead{MJD$^\tablenotemark{c}$} & \colhead{Exp.} & \colhead{S/N$^\tablenotemark{d}$} \\
\colhead{} & \colhead{(J2000)} & \colhead{(J2000)} & \colhead{} & \colhead{} & \colhead{} & \colhead{} & \colhead{}  & \multicolumn{2}{c}{($-22.5$\,mag)} & \colhead{} & \colhead{} & \colhead{} & \colhead{(s)} & \colhead{}
}\startdata
J0033.5$-$1921 & 00\,33\,34.2\, & $-$19\,21\,33\, & RBS 0076 & J0033$-$1921  & 1.67$\pm$0.11 & $-$0.597$\pm$0.010 & $>$0.505 & 0.72 & 0.42 &  1.60 & HBL & 56217 & 1950 & 290\\
J0035.2+1515 & 00\,35\,14.7\, & 15\,15\,04\, & RBS 0082 & J0035+1515 & 1.59$\pm$0.11 & $-$0.787$\pm$0.015 &  ... & 0.64 & 0.16 &  1.63 & HBL & 55775 & 910 & 150\\
J0222.6+4302 & 02\,22\,39.6\, & 43\,02\,08\, & 3C66A & J0222+4302  & 2.61$\pm$0.11 & $-$1.279$\pm$0.001 &  ... & 0.30 & 0.18 &  1.59 & IBL & 55830 & 900 & 550\\
J0316.1+0904 & 03\,16\,12.7\, & 09\,04\,43\, & RGB J0316+090 & J0316+0904 & 2.12$\pm$0.11 & $-$0.758$\pm$0.011 & ...  & 0.32 & 0.20 &  1.61 & HBL & 56217 & 1200 & 230\\
J0622.9+3326 & 06\,22\,52.2\, & 33\,26\,10\, & B2 0619+33 & J0622+3326 & 0.34$\pm$0.12 & $-$0.330$\pm$0.950 & 1.062 & ... & ... & ... & FSRQ & 55978 & 640 & 6.8\\
J1555.7+1111 & 15\,55\,43.0\, & 11\,11\,24\, & PG 1553+113 & J1555+1111  & 2.88$\pm$0.11 & $-$0.672$\pm$0.003 & ...  & 0.12 & $-$0.13 & 1.59 & HBL & 55775 & 450 & 450\\
J1559.0+5627 & 15\,58\,48.3\, & 56\,25\,14\, & 6C B155742.3+563358 & J1558+5625  & 1.49$\pm$0.11 & $-$1.174$\pm$0.001 & $>$1.049 & 0.61 & 0.26 &  1.67 & IBL & 56158 & 660 & 140\\
J1719.3+1744 & 17\,19\,13.0\, & 17\,45\,06\, & HB89 1717+178 & J1719+1745 & 0.87$\pm$0.12 & $-$1.580$\pm$0.160 &  ... & 0.75 & 0.35 & 1.80 & LBL & 56124 & 3160 & 68\\
J1754.3+3212 & 17\,54\,11.8\, & 32\,12\,23\, & RGB J1754+322 & J1754+3212  & 1.76$\pm$0.11 & $-$0.922$\pm$0.030 &  ... & 0.57 & $-$0.01 &  1.61 & HBL & 56046 & 600 & 170\\
J2039.1$-$1046 & 20\,39\,00.7\, & $-$10\,46\,42\, & CRATES J2039-1046 & J2039$-$1046 &  1.85$\pm$0.11 & $-$1.032$\pm$0.011 & ... & 0.48 & 0.14 &  1.62 & LBL & 55775 & 1000 & 240
\enddata
\label{tab:props}
\tablenotetext{a}{Measured at $\nu = 10^{14.7}$\,Hz, in units of
$10^{-28}$\,erg\,cm$^{-2}$\,s$^{-1}$\,Hz$^{-1}$.}
\tablenotetext{b}{FSRQ = flat-spectrum radio quasar; H/I/L BL = high/intermediate/low synchrotron peaked BL Lac; also called H/I/L SPs. For details, see \citet{2LAC}.}
\tablenotetext{c}{Date of Keck observation.}
\tablenotetext{d}{S/N per resolution element measured at $5000\,$\AA.}
\end{deluxetable*}
\clearpage
\end{landscape}

\end{document}